\documentclass[aps, prl, showpacs, twocolumn, superscriptaddress, floatfix,]{revtex4}
\usepackage{amsmath,amssymb,graphicx,color,overpic,float}

\begin{document}

\newcommand{\xim}{\xi_1}
\newcommand{\xic}{\xi_2}

\title{Hidden Criticality of Counterion Condensation Near a Charged Cylinder}

\author{Minryeong Cha}
\affiliation{Graduate School of Nanoscience and Technology, 
Korea Advanced Institute of Science and Technology, Deajeon 34141, Korea}
\author{Juyeon Yi}
\email{jyi@pusan.ac.kr}
\affiliation{Department of Physics, Pusan National University,
Busan 46241, Korea}
\author{Yong Woon Kim} 
\email{y.w.kim@kaist.ac.kr}
\affiliation{Graduate School of Nanoscience and Technology, 
Korea Advanced Institute of Science and Technology, Deajeon 34141, Korea}

\date{\today}

\hspace{2cm} \\

\begin{abstract}
We study the condensation transition of counterions on a charged cylinder via Monte Carlo simulations.
Varying the cylinder radius systematically in relation to the system size, we find that all counterions are bound to the cylinder and the heat capacity shows a drop at a finite Manning parameter. 
A finite-size scaling analysis is carried out to confirm the criticality of the complete condensation transition, yielding the same critical exponents with the Manning transition.
We show that the existence of the complete condensation is essential to explain how the condensation nature alters from continuous to discontinuous transition.
\end{abstract}

\maketitle

Counterion condensation is of fundamental importance in understanding static and dynamic properties of various charged polymers in ionic solutions~\cite{Bloomfield}.
When a charged polymer is very stiff, it can be considered as an infinite charged cylinder in the presence of neutralizing counterions that are confined in a cylindrical cell~\cite{Fuoss}.
A logarithmic electrostatic potential due to an oppositely charged cylinder attracts counterions and competes with the ion confinement entropy also having the similar system size dependence.
As a consequence, a characteristic phase transition occurs in the infinite dilution limit~\cite{Manning, Oosawa}; above a critical line charge density (or equivalently, below a critical temperature), 
a finite fraction of counterions remains bound close to the charged cylinder even in the limit of infinite system size~\cite{LeBret}.
This problem initially posed by Onsager and later elaborated by Manning~\cite{Manning} and Oosawa~\cite{Oosawa},
is now well recognized as a vital example to exhibit a critical phenomenon~\cite{Andelman}.  

As a representative feature of the Coulomb fluids, and also as a special case of the Kosterlitz-Thouless transition~\cite{Kholodenko},
the counterion condensation is a problem of considerable significance in theoretical physics, not to mention its importance in polymer science.
Extensive analytic~\cite{Deshkovski} and numerical~\cite{Deserno} works have been performed to study the various effects such as fluctuations and correlations~\cite{Orland, Burak, Naji}, neglected in the mean-field (MF) Poisson-Boltzmann (PB) equation.
For example, through Monte Carlo (MC) simulations, Naji and Netz found that the ionic correlations do not affect the critical temperature predicted by MF theory and the critical exponents appear to be universal, i.e., independent of correlation strength~\cite{Naji}.

Despite the fact that the counterion condensation has been studied in great detail for a very long time,
the following important questions remain to be addressed. According to the strong-coupling (SC) theory valid at asymptotically infinite ionic correlations,
the condensed fraction of counterions should abruptly change from zero to one at the critical point, suggesting a discontinuous transition. 
In contrast, MF theory always predicts a continuous transition at the same critical point. 
The first question is then, (i) how can the continuous transition in MF prediction alter its nature into the discontinuous transition in the SC theory? 
One may conjecture that the transition nature is modified by increasing the coupling strength.
However, the criticality of the Manning transition was shown to be in complete accord with MF theory even at considerably large coupling strengths~\cite{Naji}.
This gives rise to the second question, (ii) why is the criticality of the system independent of the coupling strength?

In this Letter, we aim to answer the questions using MC simulations.
One of our main findings is that in addition to the well-known Manning transition, there exists another continuous transition into a completely condensed phase.
A finite-size scaling analysis demonstrates that the complete condensation is a critical phenomenon with scale-invariant properties.
We find that the existence of the complete condensation is the key to reconcile the continuous condensation transition of MF theory with the discontinuous transition of the SC theory:
In the SC limit, the transition point of the complete condensation merges into the Manning transition point to show discontinuous transition.
If the two critical points are separated from each other, the Manning transition maintains its criticality predicted by MF theory even at elevated coupling strengths.

We consider a charged cylinder of radius $R$ and surface charge density $\sigma$ (or equivalently, line charge density $\tau$), which is placed at the origin and confined inside a cylindrical cell of radius $D$. The system is salt-free but contains $N$ neutralizing pointlike counterions of valency $q$ that are distributed between the two concentric cylinders, $r\in (R, D)$ with $r$ being the radial coordinate of counterion. The system is held at thermal equilibrium with temperature $T$. The Hamiltonian of the system reads (in units of $k_BT$) 
$
\mathcal{H}
= q^{2}\ell_{B}\sum_{\langle ij \rangle} 1/|\boldsymbol{\mathrm{x}}_{i}-\boldsymbol{\mathrm{x}}_{j}| + 2\xi\sum_{i=1}^{N} \mathrm{ln} (r_{i}/R)
$~\cite{Naji},
where $\ell_B = e^2 /\epsilon k_BT$ is the Bjerrum length. 
The Manning parameter, $\xi = q \ell_B \tau$, corresponds to the rescaled line charge density (or equivalently, to the rescaled inverse temperature)
and measures the attractive potential strength by the bare cylinder.
To minimize artificial finite size effects, we perform MC simulations with periodic boundary conditions along the cylinder axis ($z$ direction)~\cite{Naji}. 
The lateral extension parameter, $\Delta = \ln (D/R)$, is taken to be a large value to examine the criticality of the condensation phenomena in the infinite dilution limit ($\Delta \rightarrow \infty$).
For an efficient phase space sampling, we employ not only the centrifugal sampling~\cite{Naji} but also a simple global move interchanging condensed and unbound counterions, which is important in achieving an equilibration process that is independent of the initial conditions~\cite{Trizac}.

In Fig.~1(a), we plot the radial distribution function $\tilde \rho (r)= \rho(r)/2\pi \ell_B \sigma^2$ of counterions for various values of the coupling parameters $\Xi=2\pi q^3 \ell_B^2 \sigma$~\cite{Netz} at $\xi=2$.
For $\Xi=0.1$ and $10^{45}$, the MC results (symbols) compare well with MF theory (solid line) and SC theory (dashed line), respectively.
Our numerical result for the first time captures the SC limiting behavior in a cylindrical geometry.
It is well known that for planar geometries, the coupling parameter $\Xi$ measures deviations from MF theory~\cite{Netz, Moreira, BAO}.
However, we show in Fig.~1(b) that even at large $\Xi$ the density profiles are in good agreement with the PB results if $\Xi/\xi^{2}$ is small.
This result indicates that $\Xi/\xi^2$ plays a role as the effective coupling parameter in cylindrical geometry, which can be supported by a simple argument. 
For a charged cylinder, the length scale for the thickness of the condensed counterion layer (i.e., the radial extension of condensed density profile) is given by $R$~\cite{Trizac} (see also Fig.~1 (a)).
As lengths are rescaled by the thickness of the electric double layer, the ratio between the prefactors of the first and the second terms in $\mathcal{H}$ is $\Xi/\xi^2$,
which determines the relative importance of ion-ion repulsion to ion-cylinder attraction~\cite{planar}.
The effective coupling strength can be converted to the rescaled inverse radius of the cylinder, i.e., $\Xi/\xi^2=a/R$ with $a=q/\tau$ proportional to the inverse line charge density.
The relevance of $\Xi/\xi^2$ as a key parameter characterizing different regimes (e.g., thick and thin cylinder cases) has also been addressed in other literatures~\cite{Trizac, Kanduc}.

In order to identify characteristic phases, we consider the condensed fraction $f$ as an order parameter,
which is defined as the number of counterions residing in a region, $R\leq r \leq r_\ast$:
$
f=(2 \pi q/\tau)\int_{R}^{r_\ast} dr'~r' \rho(r').
$
Here, $r_\ast$ is the inflection point of the accumulated density profile, 
which is approximated in this work as $\ln (r_\ast/R)=\Delta /2$~\cite{Naji}.
According to PB equation, a finite fraction of the counterions is condensed if $\xi > 1$, as displayed by the condensed fraction in Fig.~1(c). 
On the other hand, the SC theory predicts two phases, either all counterions are condensed onto the cylinder or totally decondensed [Fig.~1(d)]. 
The transition occurs also at $\xi=1$, but it is a discontinuous phase transition in contrast to the Manning transition. 

\begin{figure}[t]
\center
\includegraphics[width=1.0\linewidth]{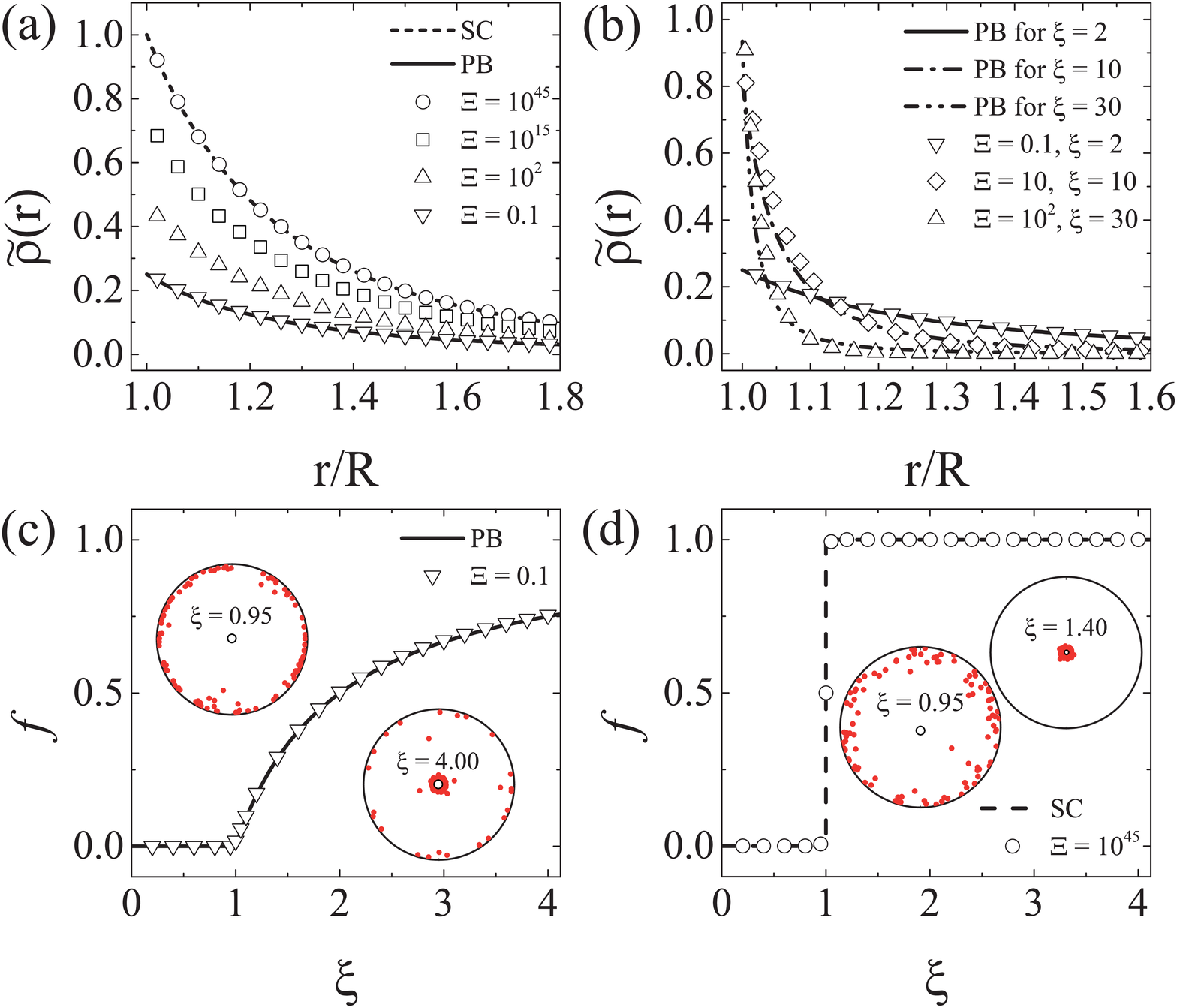}
\caption{
(a) Density profiles of counterions for various coupling parameters $\Xi$, are obtained at the Manning parameter $\xi=2$ as a function of the radial distance from the charged cylinder axis. 
(b) Density profiles for greater $\xi$ are well described by the PB predictions when $\Xi/\xi^2 \sim 0.1$.
The lower two panels display the condensed fractions as a function of $\xi$ (c) for a weakly coupled system ($\Xi =0.1$) and (d) for a strongly coupled system ($\Xi=10^{45}$).  MC simulations are performed for a given lateral parameter $\Delta=100$ with $N=100$ counterions. For both cases (c) and (d), the simulation data (symbols) are well fitted by the PB and SC theory obtained in the limit $\Delta \rightarrow \infty$. The circular images are the top views of simulation snapshots of ion distribution at the noted values of $\xi$.
}
\end{figure}

The criticality of the counterion condensation is examined in the limit of $\Delta \rightarrow \infty$,
which is indeed realized by an infinite dilution limit ($D\rightarrow \infty$) and gives $\xim=1$ as the critical fixed point of the Manning transition.
The condition $\Delta\rightarrow \infty$ is also met for the line charge limit ($R\rightarrow 0$).
Since only the ratio $D/R$ enters as a relevant parameter characterizing the system size in both MF and SC theories,
the equivalence of the two limits has often been assumed for the condensation phenomenon.
However, as shown in Fig. 1(b), $a/R$ determines the effective coupling strength of the system, and
the line charge limit is not fully equivalent to the infinite dilution limit.
In order to systematically compare the two limits, we propose a scaling relation, $R/a \sim (D/a)^{-\alpha}$ so that the two limits can be approached with different rates by controlling $\alpha$.
The advantage of employing such a relation becomes more transparent in the following analysis.

In Fig.~2(a), we plot the order parameter $f$ with increasing $\alpha$ while fixing the lateral extension parameter large as $\Delta = 100$,
which corresponds to reducing $R$ according to $\ln R/a = -\alpha \Delta /(\alpha +1)$.
As $\xi$ varies, the condensed fraction in the considered cases shows three characteristic phases, decondensation ($f=0$), partial condensation ($0 < f <1$), and complete condensation ($f=1$).
All counterions are unbound, i.e., $f=0$, for $\xi <1$, consistent with previous studies, but the complete condensation takes place at $\xi> \xic$ with $\xic$ depending on the exponent $\alpha$ roughly as 
\begin{equation}\label{secondcp}
\xic \approx (1+\alpha)/\alpha~.
\end{equation} 
To discuss the condensation in terms of thermodynamic quantities, we also measure the singular behavior of the dimensionless heat capacity per particle $\tilde{C}= \langle (\mathcal{\delta H})^2 \rangle /N$, where $\mathcal{\delta H} = \mathcal{H} - \langle \mathcal{H} \rangle$ [Fig.~2(b)].
The existence of ordered phases where all counterions reside
either in the outer region $r>r_\ast$ (decondensed phase) or in the inner region $r< r_\ast$ (completely condensed phase) is well reflected by $\tilde{C}\approx 0$ either for $\xi < \xi_1$ or for $\xi > \xic$.  
The heat capacity jumps at $\xim=1$, signaling the continuous phase transition, and 
then abruptly drops at $\xic$, indicating that the system has dual critical points of continuous phase transition. 
The locations of the jump and the drop in the heat capacity coincide with the onset points of $f=0$ and $f=1$, respectively. 

\begin{figure}[t]
\center
\includegraphics[width=1.0\linewidth]{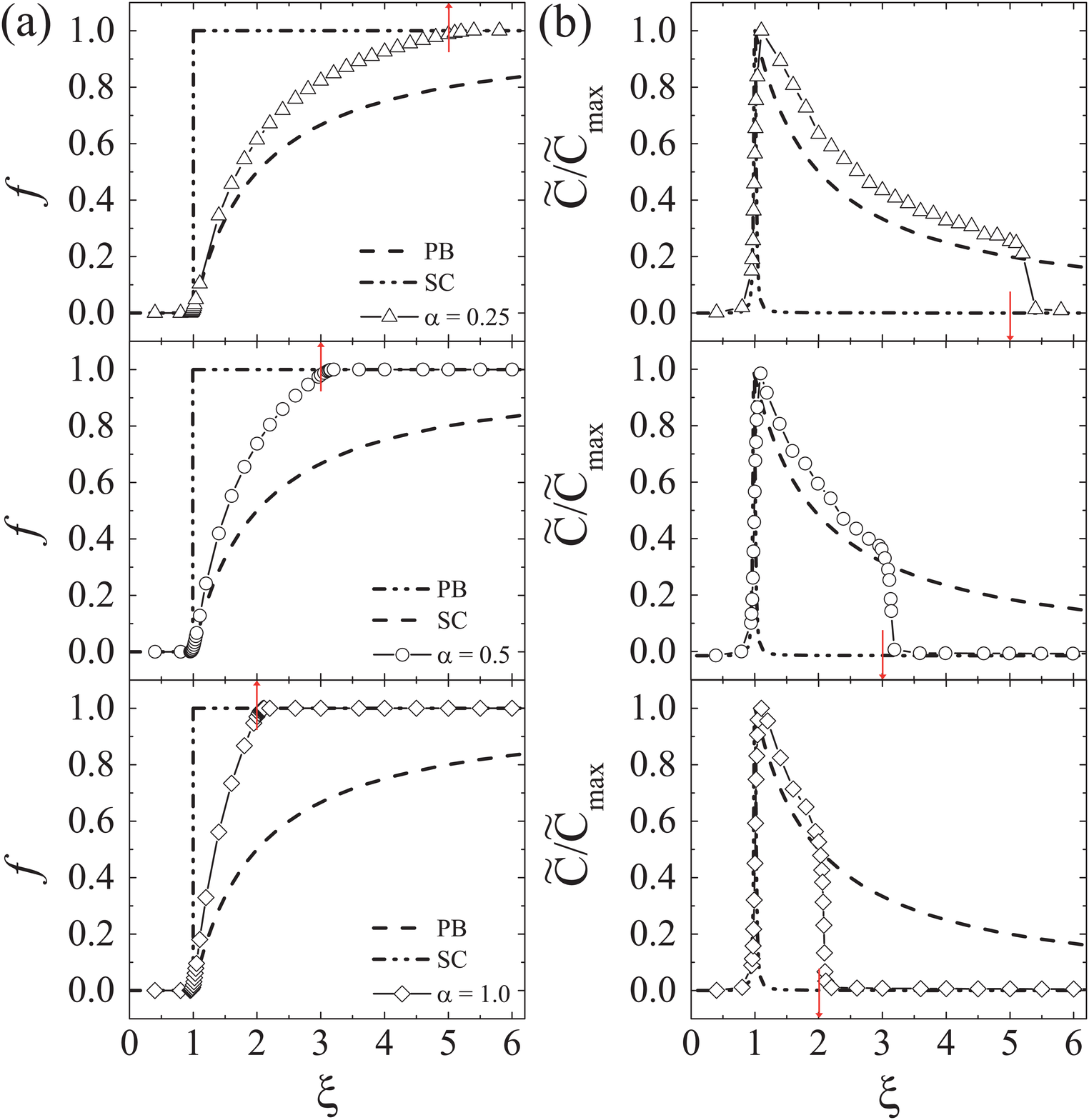}
\caption{
(a) Condensed counterion fraction and (b) heat capacity are obtained as a function of the Manning parameter $\xi$ for various $\alpha$,
where the heat capacity is scaled by its maximum value, ${\widetilde {C}}_{\mbox{max}}$. 
Simulations are performed with $\Delta = 100$ and $N=300$ counterions. 
The marked values by arrows are $\xi=(1+\alpha)/\alpha$ for each given $\alpha$, at which the condensed fraction is approximately unity, and 
the heat capacity drops down to zero. The curves are presented for comparison with the PB and SC theories. 
As $\alpha$ increases, the onset point of complete condensation approaches $\xi=1$, signaling a discontinuous transition in the limit of $\alpha \rightarrow \infty$.
}
\end{figure}

The location of $\xic$ and the presence of the complete condensation can be understood as follows:
Suppose that $N$ counterions are all bound and form a quasi one-dimensional lattice on the cylinder in a completely condensed phase.
As one of $N$ counterions diffuses to the outer boundary, the free energy difference (per simulation box) from the previous configuration is roughly estimated as
\begin{equation}\label{encri}
\delta F \approx 2(\xi-1)\ln (D/R) -2\xi \ln (D/a)~.
\end{equation}
The first term comes from a potential energy by the bare cylinder and a configurational entropy,
and the second term is the energy difference in the Coulomb repulsion with other counterions.
In the thermodynamic limit ($\Delta \rightarrow \infty$), the onset point of the complete condensation is determined by $(\delta F)_{\xi=\xic}=0$,
which leads to $\xic$ given by Eq.~(\ref{secondcp}) for $R/a\sim (D/a)^{-\alpha}$.
It follows from Eq.~(\ref{secondcp}) that for a finite $\alpha$ ($0 < \alpha < \infty$) where $R \rightarrow 0$ with $D \rightarrow \infty$,
the complete condensation transition occurs at a finite $\xic$ as re-entering the ordered phase.
For a vanishingly small $\alpha$, i.e., $\alpha \rightarrow 0$, $R$ remains finite even in the limit of $D \rightarrow \infty$, and the second transition point becomes $\xic \rightarrow \infty$. 
This indicates that the complete condensation phase for a finite radius (or a finite coupling strength) exists in the limit of $\xi \rightarrow \infty$, or equivalently in the zero-temperature limit.  
The condensation behavior is then governed by MF theory predicting only the Manning condensation at $\xi=1$, as reflected in Fig.~1(c).
When $\alpha \rightarrow \infty$, $R \rightarrow 0$ even for a large but finite $D$, and the second transition point merges into the Manning transition point as $\xic \rightarrow \xi_1$.
The heat capacity then develops an asymptotically diverging peak at $\xi=1$, signaling a discontinuous transition in agreement with the SC theory.
The simulation results suggest that the presence of the second transition bridges the two extreme behaviors of PB and SC theories.

\begin{figure}[t]
\center
\includegraphics[width=1.0\linewidth]{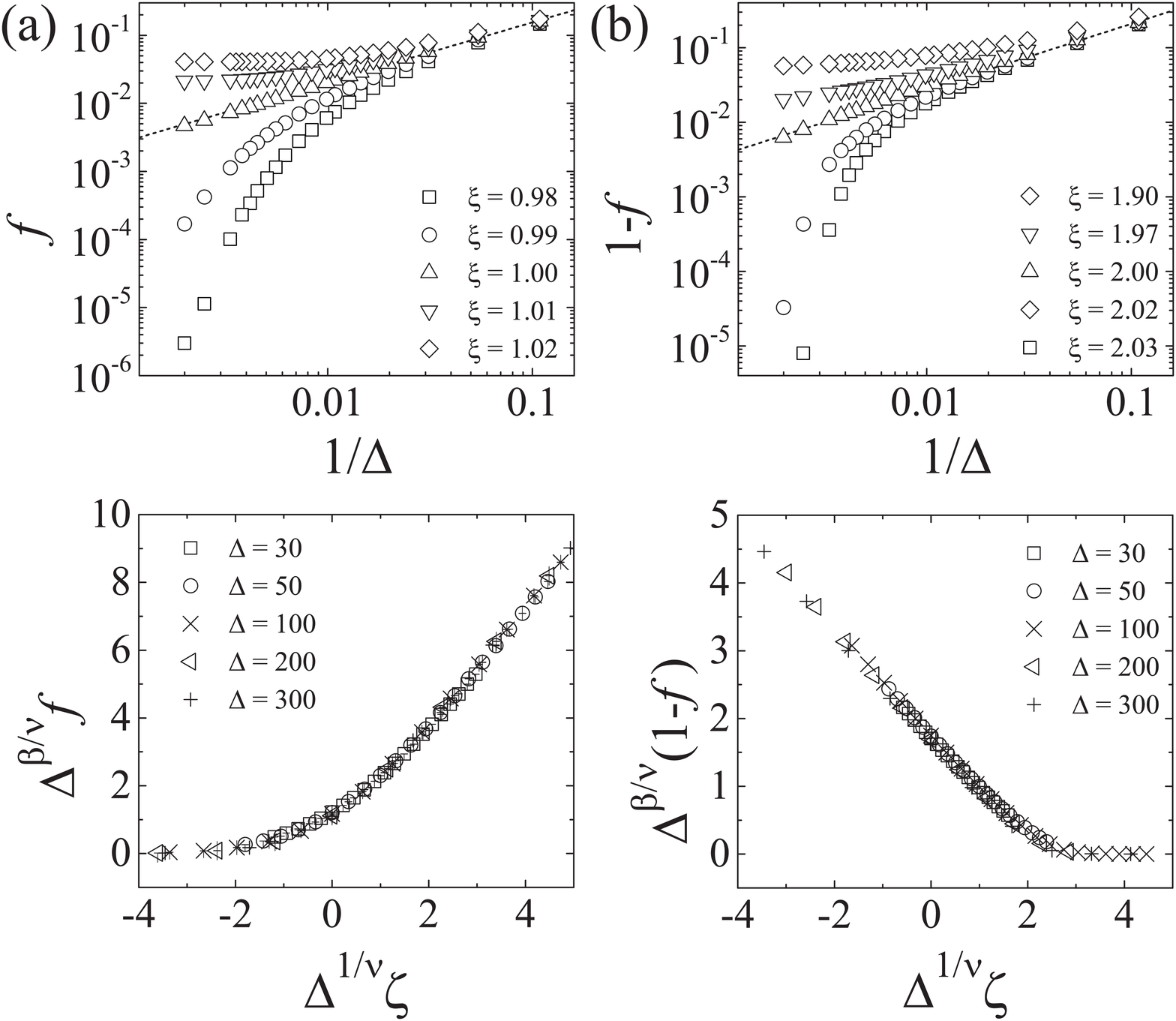}
\caption{
Scaling behaviors of the order parameter $f$ near (a) $\xi_{1}=1$ and (b) $\xi_{2}=2$ for $\alpha =1$. 
We obtain the condensed fraction, varying the lateral parameter $\Delta$ with $N=300$, and choose 
order parameters, $f\equiv m_{1}$ for the decondensation transition at $\xi=\xi_{1}$ and $1-f\equiv m_{2}$ for 
the complete condensation at $\xi=\xi_{2}$. 
The order parameters exhibit a power-law behavior near the critical points, $m_{k}\sim \Delta^{-\beta/\nu}$, where the exponent is given by
 $\beta/\nu=0.88\pm 0.01$ for both $k=1$ and $k=2$ (these scaling behaviors are presented by the dashed lines).
It is also confirmed that the order parameters satisfy a scaling relation $\Delta^{\beta/\nu} m_{k}=\widetilde{m}_{k}(\Delta^{1/\nu}\zeta_{k})$
with the reduced Manning parameter $\zeta_{k}=1-\xi_{k}/\xi$. The scaling exponents are estimated as $\beta=0.98$ and $\nu=1.11$ for both $k=1$ and $k=2$, which lead to an excellent data collapse onto a single curve.
}
\end{figure}

In order to verify the criticality from finite-sized MC simulations, we now turn to the finite-size scaling analysis of the order parameter.
We take, for example, $\alpha=1$ and perform the finite-size scaling upon varying the lateral extension parameter $\Delta$.
 In analyzing the Manning transition (near $\xi =\xim$), we choose $f \equiv m_1$ as an order parameter, and for the complete condensation transition near $\xi=\xic$, the scaling behavior of $1-f\equiv m_2$ is examined. Figure~3 shows the order parameter as a function of $1/\Delta$ for various $\xi$ values in the vicinity of the respective transition points. 
 Near $\xi=\xim$ and $\xi=\xic$, the corresponding order parameter shows a power-law decay 
 $m_{k} \sim \Delta ^{-\beta/\nu}$ with $k=1,2$, indicating a scale-invariance property. Here the exponent $\beta/\nu$ is roughly given by $\beta/\nu \approx 0.9$ both for
$m_1$ of the Manning transition near $\xi=1$ and for $m_2$ of the complete condensation transition near $\xi=2$. Also through a data collapse, we find that a scaling relation,
$m_k=\Delta^{-\beta/\nu} {\widetilde m}_{k} (\Delta^{1/\nu}\zeta_{k})$ with a scaling function ${\widetilde m}_k$ and the reduced Manning parameter $\zeta_{k}\equiv 1-\xi_{k}/\xi$, holds, giving the critical exponent $\nu\approx 1.1$ and $\beta\approx 1.0$. 
Our numerical results clearly indicate that the complete condensation around at $\xi=\xi_{2}$ is a critical phenomenon, and its critical exponents are shown identical,
within our numerical accuracy, with those of the Manning transition at $\xi=1$.  

It would be challenging to systematically vary $R/a$ and to realize the line charge limit in experiments,
but our study is of considerable theoretical interest to reveal the hidden criticality of counterion condensations that has been studied for several decades.
Furthermore, introducing the effective coupling parameter as inverse radius of the cylinder, we have recovered the MF and SC behaviors numerically.
We have also shown that the presence of complete condensation is a key to connect the two limiting theories,
i.e., how continuous transition of condensation predicted by the MF theory gradually evolves into discontinuous transition by the SC theory.
It would be interesting to extend the present study to further explore the effect of internal structures of counterions on the condensation transitions~\cite{Kim}.


\begin{thebibliography}{10}
\bibitem{planar}
For planar geometries, the thickness of the condensed counterion layer is characterized by the Gouy-Chapman length, $\mu=1/2\pi q\ell_B \sigma $~\cite{Andelman2}.
As rescaling the Hamiltonian for a system with a charged plane by $\mu$, one thus finds that 
the relative strength of ion-ion repulsion to ion-plate attraction is given by the well-known coupling parameter $\Xi=2\pi q^3 \ell_B^2 \sigma$
in case of planar geometries~\cite{Netz, Moreira, BAO}.
\end{thebibliography}
\end{document}